\documentclass[twocolumn,showpacs,preprintnumbers,amsmath,amssymb,pra]{revtex4}

\usepackage{graphicx}
\usepackage{dcolumn}
\usepackage{bm}

\begin{document}


\title{Division of the momentum of electromagnetic waves in linear media into electromagnetic and material parts}

\author{Pablo L. Saldanha}
 \email{pablols@fisica.ufmg.br}
\affiliation{%
Departamento de F\'isica, Universidade Federal de Minas Gerais,
Caixa Postal 702, 30123-970, Belo Horizonte, Minas Gerais, Brazil
}%


\begin{abstract} It is proposed a natural and consistent division of
the momentum of electromagnetic waves in linear, non-dispersive and
non-absorptive dielectric and magnetic media into material and
electromagnetic parts. The material part is calculated using
directly the Lorentz force law and the electromagnetic momentum
density has the form $\varepsilon_0\mathbf{E}\times\mathbf{B}$,
without an explicit dependence on the properties of the media. The
consistency of the treatment is verified through the obtention of a
correct momentum balance equation in many examples and showing the
compatibility of the division with the Einstein's theory of
relativity by the use of a \emph{gedanken} experiment. An
experimental prediction for the radiation pressure on mirrors
immersed in linear dielectric and magnetic media is also made.
\end{abstract}

\pacs{03.50.De, 41.20.Jb}
\maketitle

\section{Introduction}

There has been an extensive debate about the correct expression for
the momentum density of electromagnetic waves in linear media. The
Minkowski's expression $\mathbf{D}\times \mathbf{B}$ and the
Abraham's $\mathbf{E}\times \mathbf{H}/c^2$ are the most famous
ones, both proposed in the beginning of the twentieth century
\cite{pfeifer07}. In these expressions, $\mathbf{E}$ is the electric
field, $\mathbf{B}$ is the magnetic field, $\mathbf{D}\equiv
\varepsilon_0\mathbf{E}+\mathbf{P}$ is the electric displacement
field and $\mathbf{H}$ is defined as $\mathbf{H}\equiv
\mathbf{B}/\mu_0 -\mathbf{M}$, where $\mathbf{P}$ and $\mathbf{M}$
are the electric and magnetic dipole densities of the medium.
$\varepsilon_0$ is the permittivity of free space, $\mu_0$ is the
permeability of free space and $c={1}/{\sqrt{\varepsilon_0\mu_0}}$
is the speed of light in vacuum. These expressions make
qualitatively different predictions for the momentum of light in a
medium. The Minkowski formulation predicts that a photon with
momentum ${\hbar\omega}/{c}$ in vacuum increases its momentum to
$n{\hbar\omega}/{c}$, where $n$ is the refraction index of the
medium, on entering a dielectric medium. The Abraham expression,
however, states that the photon momentum decreases to
${\hbar\omega}/{nc}$ on entering the medium.

The debate of which form is the correct one persisted for many
decades, with experiments and theoretical discussions, from time to
time, seeming to favor either one of the two formulations. Arguments
that sometimes are naively used in favor of the Minkowski
formulation are the experiments of Jones \emph{et al.}
\cite{jones54,jones78} that measured the radiation pressure on
mirrors immersed in dielectric media, the experiments of Ashkin and
Dziedzic \cite{ashkin73} that measured the radiation pressure on the
free surface of a liquid dielectric, the experiments of Gibson
\emph{et al.} \cite{gibson80} that measured the radiation pressure
on the charges of a semiconductor via the photon drag effect and the
experiments of Campbell \emph{et al.} \cite{campbell05} that
measured the recoil momentum of atoms in a gas after absorbing one
photon. All these experiments are consistent with the consideration
that a photon in a medium has momentum $n{\hbar\omega}/{c}$.
Arguments that sometimes are naively used in favor of the Abraham
formulation are the symmetry of its energy-momentum tensor,
compatible with conservation of angular momentum, the agreement of
its electromagnetic momentum density with the predictions of
Einstein box theories \cite{frisch65,brevik81,loudon04} in a direct
way, the experiments of Walker \emph{et al.}
\cite{walker75b,walker77} that measured the torque on a dielectric
disk suspended in a torsion fiber subjected to external magnetic and
electric fields and the experiments of She \emph{et al.}
\cite{she08} that measured a push force on the end face of a silica
filament when a light pulse exits it. These experiments are
consistent with the Abraham form for the momentum density of an
electromagnetic wave in a dielectric medium.

Although the debate is still supported by some researchers, Penfield
and Haus showed, more than forty years ago, that neither of the
forms is completely correct on its own \cite{penfield}.  The
electromagnetic momentum of electromagnetic waves in linear media is
always accompanied by a material momentum, and when the material
counterpart is taken into account, both the Minkowski and Abraham
forms for the electromagnetic momentum density yield the same
experimental predictions.  They are simply two different ways, among
many others, to divide the total momentum density.  A revision of
this discussion and the eventual conclusion can be found in Ref.
\cite{pfeifer07}.  In fact, the experimental results of Jones
\emph{et al.} were reproduced by Gordon \cite{gordon73} and Loudon
\cite{loudon02} using the Abraham form for the electromagnetic
momentum density and calculating the material momentum by means of
the Lorentz force.  Gordon also reproduced the results of Ashkin and
Dziedzic by the same way \cite{gordon73}.  Loudon \emph{et al.}
\cite{loudon05} showed that the experiments of Gibson \emph{et al.}
can also be explained with both formulations.  And Leonhardt
\cite{leonhardt06} showed that the experiments of Campbell \emph{et
al.} can also be explained by the use of the Abraham form for the
momentum density and a redefinition of the mechanical momentum.  On
the other side, Israel \cite{israel77} has derived the experimental
results of Walker \emph{et al.} using the Minkowski formulation with
a suitable combination of electromagnetic and material
energy-momentum tensors and the conclusions of the experiments of
She \emph{et al.} were recently questioned
\cite{mansuripur09c,mansuripur09a,brevik09}.

What happens is that for each experimental situation one formulation
can predict the behavior of the system in a simpler way, but the
Minkowski, Abraham and other proposed formulations are always
equivalent. In a recent article, Pfeifer \emph{et al.} show the
conditions in which we can neglect the material counterpart of the
Minkowski energy-momentum tensor \cite{pfeifer09}, justifying why it
is possible to predict the behavior of the experiments of Jones
\emph{et al.} and the modeling of optical tweezers only with the
electromagnetic tensor in the Minkowski formulation. To summarize, I
quote Ref. \cite{pfeifer07}: ``(...) no electromagnetic wave
energy-momentum tensor is complete on its own. When the appropriate
accompanying energy-momentum tensor for the material medium is also
considered, experimental predictions of all various proposed tensors
will always be the same, and the preferred form is therefore
effectively a matter of personal choice.''

But we can ask if there is, among all possible ways to divide the
total momentum density of an electromagnetic wave in a medium into
electromagnetic and material parts, a natural one. I believe there
is. $\mathbf{E}$ and $\mathbf{B}$ are the fields that appear in the
Lorentz force law and actually interact with electric charges. They
are the fields that can transfer momentum to matter. So, in my point
of view, they must be the fields that may carry electromagnetic
momentum. $\mathbf{D}$ and $\mathbf{H}$ should be seen as averaged
quantities of material and electromagnetic properties, used to
simplify the calculations. In this sense, it seems natural that the
electromagnetic part of the momentum density of an electromagnetic
wave in a medium has the form
$\mathbf{p}_{\mathrm{e.m.}}=\varepsilon_0\mathbf{E}\times\mathbf{B}$,
which does not have an explicit dependence on the properties of the
medium. The material part of the momentum should be calculated as
the momentum acquired by the medium by the action of the Lorentz
force on the charges of the medium. An electromagnetic
energy-momentum tensor that has these characteristics was previously
proposed by Obukhov and Hehl \cite{obukhov03}. Here I show the
validity of this division of the momentum density in a series of
examples. The form $\varepsilon_0\mathbf{E}\times\mathbf{B}$ for the
electromagnetic momentum density is equivalent to the Abraham one in
non-magnetic media. As in all cited experiments the media under
consideration were non-magnetic, there is no difference between the
treatment with this momentum density and the Abraham one. Only in
magnetic media the differences will appear.

Gordon \cite{gordon73}, Loudon \cite{loudon02,loudon03} and
Mansuripur \cite{mansuripur04} calculated the material momentum
density of electromagnetic waves in linear non-dispersive
dielectrics using directly the Lorentz force law. Scalora \emph{et
al.} \cite{scalora06} used numerical simulations to calculate, also
by the Lorentz force law, the momentum transfer to more general
dispersive media. Recently, Hinds and Barnett \cite{hinds09} used
the Lorentz force to calculate the momentum transfer to a single
atom. Here, following this method in an analytical treatment, I show
that it may exist permanent transfers of momentum to the media due
to the passage of electromagnetic pulses that were not considered
before. I also generalize the previous treatments considering
magnetic media. Mansuripur has treated magnetic materials in another
work \cite{mansuripur07b}, but the force equation, results and
conclusions of this work are different from his. I believe that my
treatment is more adequate. If this method of using the Lorentz
force is adopted to calculate the material momentum of
electromagnetic waves in linear media, we conclude that the
electromagnetic part of the momentum density must have the form
$\varepsilon_0\mathbf{E}\times\mathbf{B}$ in order that we have
momentum conservation in the various circumstances that are
discussed in this paper.

In Sec. \ref{sec:mom}, I calculate the material momentum density of
an electromagnetic pulse in a homogeneous linear dielectric and
magnetic medium. In Sec. \ref{sec:int diel}, I calculate the
momentum transfer to the medium near the interface on the partial
reflection and transmission of an electromagnetic pulse by the
interface between two linear media and show the momentum
conservation in the process. In Sec. \ref{sec:jones}, I show the
compatibility of the present treatment with the experiments of Jones
\emph{et al.} \cite{jones54,jones78} for the momentum transfer from
an electromagnetic wave in a dielectric medium to a mirror upon
reflection, show the momentum conservation in this process and
generalize the treatment of Mansuripur \cite{mansuripur07a} for the
radiation pressure on mirrors immersed in linear media for arbitrary
kinds of mirrors and magnetic media. In Sec. \ref{sec:antirref}, I
use my method to calculate the momentum transfer to an
antireflection layer between two linear media on the passage of an
electromagnetic pulse and show the momentum conservation in the
process. In Sec. \ref{sec:einstein}, I show the compatibility of the
proposed division of the momentum density with the Einstein's theory
of relativity by the use of a \emph{gedanken} experiment of the kind
``Einstein box theories''. Finally, in Sec. \ref{conc}, I present my
concluding remarks.

\section{Material momentum of electromagnetic waves in linear
dielectric and magnetic media}\label{sec:mom}

The momentum transfer to a linear non-absorptive and non-dispersive
dielectric medium due to the presence of an electromagnetic wave can
be calculated directly using the Lorentz force
\cite{gordon73,loudon02,loudon03,mansuripur04,scalora06}. The force
acting on electric dipoles can be written as
\begin{equation}\label{lor force dipoles}
    \mathbf{F}_{\mathrm{dip.}}=(\mathbf{p}\cdot \nabla)\mathbf{E}+\frac{\mathrm{d}\mathbf{p}}{\mathrm{d}t}\times
    \mathbf{B}\;,
\end{equation} where $\mathbf{p}$ is the dipole moment. In a linear,
isotropic and non-dispersive dielectric, the electric dipole moment
density can be written as
$\mathbf{P}=\chi_\mathrm{e}\varepsilon_0\mathbf{E}$, where
$\chi_\mathrm{e}$ is the electric susceptibility of the medium. It
is important to stress that the consideration of a non-dispersive
linear medium must be seen as an approximation, once every material
medium is inevitably accompanied of dispersion. But if the
electromagnetic wave has a narrow frequency spectrum and the
dispersion is small in this region of frequencies, this treatment
will give the material contribution of the total momentum of the
wave with a good precision. Using Eq. (\ref{lor force dipoles}), the
Maxwell equations and some vectorial identities,  the force density
on this medium can be written as \cite{gordon73}
\begin{equation}\label{force density dipoles 2}
    \mathbf{f}_{\mathrm{diel.}}=\chi_\mathrm{e}\varepsilon_0\left[\nabla\left(\frac{1}{2}E^2\right)+\frac{\partial}{\partial
    t}\left(\mathbf{E}\times \mathbf{B}\right)\right]\;.
\end{equation}

In magnetized media, there is also a bound current
$\nabla\times\mathbf{M}$ that is affected by the Lorentz force. So
the force density in a linear non-dispersive dielectric and magnetic
medium can be written as

\begin{equation}\label{force density}
    \mathbf{f}=\underbrace{\chi_\mathrm{e}\varepsilon_0\nabla\left(\frac{1}{2}E^2\right)}_{\mathbf{f}_1}+\underbrace{\chi_\mathrm{e}\varepsilon_0\frac{\partial}{\partial
    t}\left(\mathbf{E}\times \mathbf{B}\right)}_{\mathbf{f}_2}+\underbrace{(\nabla\times\mathbf{M})\times\mathbf{B}}_{\mathbf{f}_3}\;.
\end{equation} I will calculate the material momentum due to the
action of the forces $\mathbf{f}_1$, $\mathbf{f}_2$ and
$\mathbf{f}_3$ separately. In a linear, isotropic and non-dispersive
magnetic medium, we have
$\mathbf{M}=\chi_\mathrm{m}\mathbf{H}={\chi_\mathrm{m}}/[{(1+\chi_\mathrm{m})\mu_0]}\mathbf{B}$,
where $\chi_\mathrm{m}$ is the magnetic susceptibility of the
medium.

In his treatment of the material part of the momentum of
electromagnetic waves in magnetic materials \cite{mansuripur07b},
Mansuripur uses a specific model for a magnetic medium, obtaining a
different equation for the bound currents in the medium. The form
$\nabla\times\mathbf{M}$ is more general and agrees with his form in
a homogeneous medium. As the interfaces between different linear
media will be treated here, the general form for the bound currents
must be used. He also takes the vector product of the bound currents
with $\mu_0\mathbf{H}$ instead of $\mathbf{B}$ to find the force
density. I don't think this is adequate. For these reasons, I
believe that the present treatment to find the material momentum of
electromagnetic waves in magnetic media is more adequate than that
of Mansuripur.

Consider an electromagnetic plane wave propagating in
$\mathbf{\hat{z}}$ direction described by the following electric
field: \begin{equation}\label{pulso}
    \mathbf{E}_\mathrm{i}(z,t)=\frac{1}{\sqrt{2\pi}}\int_{-\infty}^{+\infty} \mathrm{d}\omega \tilde{E}(\omega)\mathrm{e}^{{i\left(\frac{n\omega}{c}z-\omega
    t\right)}}\mathbf{\hat{x}}\;,
\end{equation} with $\tilde{E}(-\omega)=\tilde{E}^*(\omega)$ and
$\mathbf{B}_\mathrm{i}=({n}/{c})|\mathbf{E}_\mathrm{i}|\mathbf{\hat{y}}$,
$n=\sqrt{(1+\chi_\mathrm{e})(1+\chi_\mathrm{m})}$ being the
refraction index of the medium. The consideration of a plane wave
pulse must also be seen as an approximation. We can consider a beam
with a small angular spread, in which all wavevectors that compose
it are very close to the $z$ axis such that their $z$ component  are
equal to their norm in a good approximation. For the wave of Eq.
(\ref{pulso}), the force densities $\mathbf{f}_1$ and $\mathbf{f}_3$
from Eq. (\ref{force density}) can be written as
\begin{eqnarray}\label{f3}
     \mathbf{f}_1&=&-\frac{\chi_\mathrm{e}\varepsilon_0}{2}\frac{\partial}{\partial
      t}(\mathbf{E}\times \mathbf{B})\;,\\
      \mathbf{f}_3&=&\frac{\chi_\mathrm{m}(1+\chi_\mathrm{e})\varepsilon_0}{2}\frac{\partial}{\partial
      t}(\mathbf{E}\times \mathbf{B})\;.
\end{eqnarray} Substituting these expressions for $\mathbf{f}_1$ and
$\mathbf{f}_3$, the material momentum density of an electromagnetic
wave in a homogeneous linear non-dispersive and non-absorptive
medium can be written as \begin{equation}\label{pmat}
    \mathbf{p}_{\mathrm{mat}}(t)=\int_{-\infty}^{t}\mathrm{d}t'\mathbf{f}(t')=\frac{(\chi_\mathrm{e}+\chi_\mathrm{m}+\chi_\mathrm{e}\chi_\mathrm{m})}{2}\varepsilon_0\mathbf{E}(t)\times\mathbf{B}(t).
\end{equation} It is important to stress that this material momentum
density propagates with the pulse and disappears after its passage
through the medium. In the remaining part of this work, it will be
considered that the total momentum density of an electromagnetic
wave in a homogeneous linear medium is given by the sum of the
material momentum density above and the electromagnetic momentum
density $\varepsilon_0\mathbf{E}\times\mathbf{B}$. By also
calculating the permanent transfers of momentum to the media by the
action of the Lorentz force in some situations, the momentum
conservation in these processes and the consistency of the proposed
division of the momentum density will be shown.

\section{Reflection and transmission of an electromagnetic pulse by
the interface between two linear media}\label{sec:int diel}

\begin{figure}
  \begin{center}
  \includegraphics[width=6cm]{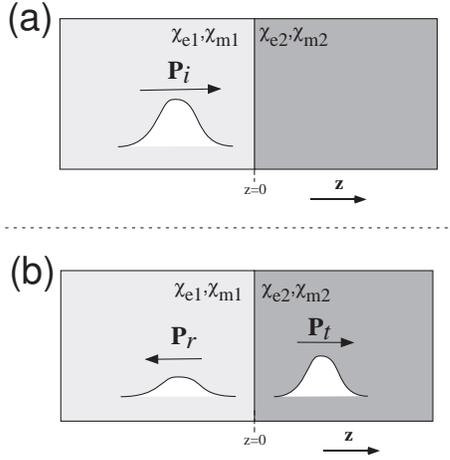}\\
  \caption{(a) A pulse with total momentum $\mathbf{P}_\mathrm{i}$ propagates in medium 1 with electric and magnetic susceptibilities
  $\chi_{\mathrm{e1}}$ and $\chi_{\mathrm{m1}}$
  towards the interface with medium 2 with electric and magnetic susceptibilities
  $\chi_{\mathrm{e2}}$ and $\chi_{\mathrm{m2}}$. (b) The resultant reflected and
  transmitted pulses with total momentum $\mathbf{P}_\mathrm{r}$ and $\mathbf{P}_\mathrm{t}$.}\label{fig-pulsos}\end{center}
\end{figure}

Consider the situation illustrated in Fig. \ref{fig-pulsos}.
Initially we have a pulse of electromagnetic radiation in medium 1,
with electric susceptibility $\chi_{\mathrm{e1}}$, magnetic
susceptibility $\chi_{\mathrm{m1}}$ and refraction index
$n_1=\sqrt{(1+\chi_{\mathrm{e1}})(1+\chi_{\mathrm{m1}})}$,
propagating in $\mathbf{\hat{z}}$ direction towards the interface
with medium 2, with electric and magnetic susceptibilities
$\chi_{\mathrm{e2}}$ and $\chi_{\mathrm{m2}}$ and refraction index
$n_2=\sqrt{(1+\chi_{\mathrm{e2}})(1+\chi_{\mathrm{m2}})}$. The
interface is in the plane $z=0$ and the incidence is normal. Later,
we will have a transmitted pulse in medium 2 and a reflected pulse
in medium 1. It will be shown that the proposed division of the
momentum density of the pulse in electromagnetic and material parts
leads to the momentum conservation in the process. Representing the
electric field of the incident pulse as in Eq. (\ref{pulso}), the
electromagnetic part of its momentum can be written as
\begin{equation}\label{p0}
    \mathbf{P}_0=\int \mathrm{d}^3r \varepsilon_0\mathbf{E}_\mathrm{i}\times
     \mathbf{B}_\mathrm{i}=\varepsilon_0\int \mathrm{d}x\int \mathrm{d}y \int_{-\infty}^{+\infty}
     \mathrm{d}\omega|\tilde{E}(\omega)|^2\mathbf{\hat{z}}\;.
\end{equation} The integrals in $x$ and $y$ are, in principle,
infinite.  But a plane wave is always an approximation, so the
amplitude $\tilde{E}(\omega)$ must decay for large $x$ and $y$.  I
will not worry about that, only assume that the integral is finite.
Integrating also the material momentum density of Eq. (\ref{pmat}),
we find that the total momentum of the incident
($\mathbf{P}_\mathrm{i}$), reflected ($\mathbf{P}_\mathrm{r}$)  and
transmitted ($\mathbf{P}_\mathrm{t}$) pulses are
\begin{eqnarray}\label{pi pr pt}\nonumber
    &&\mathbf{P}_\mathrm{i}=\left(1+\frac{\chi_{\mathrm{e1}}+\chi_{\mathrm{m1}}+\chi_{\mathrm{e1}}\chi_{\mathrm{m1}}}{2}\right)\mathbf{P}_0\;,\;\mathbf{P}_\mathrm{r}=-|r|^2\mathbf{P}_\mathrm{i}\;,\\
    &&\mathbf{P}_\mathrm{t}=|t|^2\left(1+\frac{\chi_{\mathrm{e2}}+\chi_{\mathrm{m2}}+\chi_{\mathrm{e2}}\chi_{\mathrm{m2}}}{2}\right)\mathbf{P}_0\;,
\end{eqnarray} where $r$ and $t$ are the reflection and transmission
coefficients, respectively.

There is also a momentum transfer to medium 1 during reflection that
does not propagate with the electromagnetic pulse. We can observe in
Eq.  (\ref{force density}) that it is the total (incident plus
reflected) field that generates the force density $\mathbf{f}_1$. As
$E^2=E_\mathrm{i}^2+E_\mathrm{r}^2+2\mathbf{E}_\mathrm{i}\cdot
\mathbf{E}_\mathrm{r}$ in medium 1, we must consider the term
$\mathbf{f}_1'=\chi_{\mathrm{e1}}\varepsilon_0\nabla\left(\mathbf{E}_\mathrm{i}\cdot\mathbf{E}_\mathrm{r}\right)$.
The momentum transfer due to this term is \begin{eqnarray}\label{p
linha 1} \nonumber
\mathbf{P}_1'&=&\int_{-\infty}^{+\infty}\mathrm{d}t\int
\mathrm{d}x\int \mathrm{d}y \int_{-\infty}^0 \mathrm{d}z
     \chi_{\mathrm{e1}}\varepsilon_0\frac{\partial}{\partial
     z}\left(\mathbf{E}_\mathrm{i}\cdot\mathbf{E}_\mathrm{r}\right)\mathbf{\hat{z}}\\
     &=&r\chi_{\mathrm{e1}}\mathbf{P}_0\;,
\end{eqnarray} with $\mathbf{P}_0$ given by Eq. (\ref{p0}).

As $\mathbf{E}_\mathrm{i}\times \mathbf{B}_\mathrm{r}
+\mathbf{E}_\mathrm{r}\times \mathbf{B}_\mathrm{i}=0$, the force
density $\mathbf{f}_2$ in Eq. (\ref{force density}) does not
contribute to a permanent momentum transfer to the medium. The
permanent momentum transfer from Eq. (\ref{p linha 1}) was not
considered in the previous treatments of reflection of
electromagnetic pulses by interfaces between two dielectrics
\cite{loudon02,loudon03,mansuripur04}, so these works are compatible
with momentum conservation only when the incidence medium is vacuum
and $\mathbf{P}_1'=0$.

The force density $\mathbf{f}_3$ in Eq. (\ref{force density}) also
contributes to a permanent transfer of momentum to medium 1. We can
see that for the pulse of Eq. (\ref{pulso}) this force density can
be written as \begin{equation}\label{f3}
    \mathbf{f}_3=
    -\frac{\chi_\mathrm{m}}{\mu_0(1+\chi_\mathrm{m})}\nabla\left(\frac{B^2}{2}\right)\;.
\end{equation} Repeating the calculation of Eq. (\ref{p linha 1}),
remembering that $\mathbf{B}_\mathrm{r}=-r\mathbf{B}_\mathrm{i}$, we
can see that this force density transfers a momentum $\mathbf{P}'_3$
to the medium 1 given by \begin{equation}\label{p linha 3}
    \mathbf{P}_3'=r\chi_{\mathrm{m1}}(1+\chi_{\mathrm{e1}})\mathbf{P}_0\;.
\end{equation}

There is still another contribution to the momentum transfer to the
interface due to the discontinuity of $\mathbf{M}$ in the interface.
In an extent $\delta z$ much smaller than the wavelength of light
around $z=0$, $\mathbf{f}_3$ from Eq. (\ref{force density}) can be
written as \begin{equation}\nonumber
  \mathbf{f}_3|_{z=0} =  \frac{1}{\delta z}
  \left[\frac{\chi_{\mathrm{m2}}}{1+\chi_{\mathrm{m2}}}B_2-\frac{\chi_{\mathrm{m1}}}{1+\chi_{\mathrm{m1}}}B_1\right]\frac{(-\mathbf{\hat{x}})}{\mu_0}\times\mathbf{B}\;,
\end{equation} where $\mathbf{B}_1=\mathbf{B}_\mathrm{i}(1-r)$ is
the magnetic field just before the interface, in medium 1, and
$\mathbf{B}_2=[({1+\chi_{\mathrm{m2}}})/({1+\chi_{\mathrm{m1}}})]\mathbf{B}_1$
is the magnetic field just after the interface, in medium 2.
Integrating in this region of extent $\delta z$ and in $x$, $y$ and
$t$, we obtain the following momentum transfer to this interface:
\begin{eqnarray}\label{p linha 4}
\mathbf{P}'_4&=&\int_{-\infty}^{+\infty}\mathrm{d}t\int
\mathrm{d}x\int \mathrm{d}y \int_{-\delta z/2}^{+\delta
z/2}\mathrm{d}z \;\mathbf{f}_3\\\nonumber
&=&\frac{(\chi_{\mathrm{m1}}-\chi_{\mathrm{m2}})(1+\chi_{\mathrm{e1}})(1-r)^2}{2}\left[1+\frac{1+\chi_{\mathrm{m2}}}{1+\chi_{\mathrm{m1}}}\right]\mathbf{P}_0.
\end{eqnarray}

Using Eqs. (\ref{pi pr pt}), (\ref{p linha 1}), (\ref{p linha 3})
and (\ref{p linha 4}) and substituting the values of $r$ and $t$
\cite{jackson}, \begin{eqnarray}\nonumber
    r&=&\frac{\sqrt{(1+\chi_{\mathrm{e1}})(1+\chi_{\mathrm{m2}})}-\sqrt{(1+\chi_{\mathrm{e2}})(1+\chi_{\mathrm{m1}})}}{\sqrt{(1+\chi_{\mathrm{e1}})(1+\chi_{\mathrm{m2}})}+\sqrt{(1+\chi_{\mathrm{e2}})(1+\chi_{\mathrm{m1}})}}\;,\\\nonumber
   t&=&\frac{2\sqrt{(1+\chi_{\mathrm{e1}})(1+\chi_{\mathrm{m2}})}}{\sqrt{(1+\chi_{\mathrm{e1}})(1+\chi_{\mathrm{m2}})}+\sqrt{(1+\chi_{\mathrm{e2}})(1+\chi_{\mathrm{m1}})}}\;,
\end{eqnarray} it can be shown that \begin{equation}
    \mathbf{P}_\mathrm{r}+\mathbf{P}_\mathrm{t}+\mathbf{P}'_1+\mathbf{P}'_3+\mathbf{P}'_4=\mathbf{P}_\mathrm{i}\;.
\end{equation}

The obtention of a correct momentum balance equation shows the
consistency of the proposed division of the momentum of the wave
into material and electromagnetic parts.

\section{Radiation pressure on submerged mirrors}\label{sec:jones}

In the second example, consider that we put a very good conductor in
place of medium 2 in Fig. \ref{fig-pulsos} that reflects an
electromagnetic pulse described by Eq. (\ref{pulso}). The momentum
transfer to this mirror can be calculated using the Lorentz force
that acts on the induced currents in the mirror. The magnetic field
inside the mirror ($z>0$) can be written as \cite{jackson}
\begin{equation}
    \mathbf{B}_{\mathrm{mir}}=\frac{1}{\sqrt{2\pi}}\int_{-\infty}^{+\infty} \mathrm{d}\omega \frac{2n_1}{(1+\chi_{\mathrm{m1}})c}\tilde{E}(\omega)
    \mathrm{e}^{(-\kappa+ik)z-i\omega
    t}\mathbf{\hat{y}}\;,
\end{equation} where $k$ and $\kappa$ are real functions of
$\omega$. This form for the magnetic field guarantees the continuity
of $\mathbf{H}$ at the interface and decrease exponentially with
$z$. Disregarding the small presence of the electric field inside
the mirror that would generate absorption of electromagnetic energy,
the current density in the mirror can be written as
$\mathbf{J}_{\mathrm{mir}}=({1}/{\mu_0})\nabla \times \mathbf{B}$.
So the momentum transfer to the mirror due to the reflection of the
electromagnetic pulse is \begin{eqnarray}
   && \mathbf{P}_{\mathrm{mir}}=\int \mathrm{d}x \int \mathrm{d}y \int_0^{+\infty} \mathrm{d}z \int _{-\infty}^{+\infty}
    \mathrm{d}t\,
    \mathbf{J}_{\mathrm{mir}} \times \mathbf{B}_{\mathrm{mir}}\\\nonumber &&=\int \mathrm{d}x \int \mathrm{d}y \int_{-\infty}^{+\infty} \mathrm{d}\omega \frac{2\varepsilon_0(1+\chi_{\mathrm{e1}})}{(1+\chi_{\mathrm{m1}})}|\tilde{E}(\omega)|^2\frac{(\kappa-ik)}{\kappa}\mathbf{\hat{z}}\;.
\end{eqnarray} We have $|\tilde{E}(-\omega)|=|\tilde{E}(\omega)|$,
$\kappa(-\omega)=\kappa(\omega)$ and $k(-\omega)=-k(\omega)$, such
that the integral in $\omega$ of the term that multiplies ($-ik$) is
zero. So the momentum transfer from the pulse to the mirror upon
reflection is \begin{equation}\label{mom esp}
    \mathbf{P}_{\mathrm{mir}}=2\left(\frac{1+\chi_{\mathrm{e1}}}{1+\chi_{\mathrm{m1}}}\right)\mathbf{P}_0\;,
\end{equation} with $P_0$ given by Eq. (\ref{p0}).

The same momentum transfer is obtained with the condition
$\mathbf{P}_{\mathrm{mir}}=\mathbf{P}_{\mathrm{i}}-\mathbf{P}_\mathrm{r}-\mathbf{P}'_1-\mathbf{P}'_3-\mathbf{P}'_4$
using Eqs. (\ref{pi pr pt}), (\ref{p linha 1}), (\ref{p linha 3})
and (\ref{p linha 4}) with $r=-1$ and $\chi_{\mathrm{m2}}=0$. On
this way, we arrive at the same expression (\ref{mom esp}) for
$\mathbf{P}_{\mathrm{mir}}$, showing again the consistency of the
treatment.

For non-dispersive linear media, the total energy density of the
wave can be written as
$u_{\mathrm{tot}}=(1+\chi_{\mathrm{e1}})\varepsilon_0|\mathbf{E}|^2$
\cite{jackson}. So the energy of the incident pulse of Eq.
(\ref{pulso}) is \begin{equation}\label{en pulso i}
    U_\mathrm{i}=\int \mathrm{d}^3r
    (1+\chi_{\mathrm{e1}})\varepsilon_0E_\mathrm{i}^2=\frac{(1+\chi_{\mathrm{e1}})c}{n_1}\,|\mathbf{P}_0|.
\end{equation}

The ratio between the modulus of the momentum transfer to the mirror
(\ref{mom esp}) and the incident energy (\ref{en pulso i}) is
${2n_1}/[{(1+\chi_{\mathrm{m1}})c}]$, in accordance with the
experiments of Jones \emph{et al.} \cite{jones54,jones78}. In these
experiments, the magnetic susceptibilities of the media were too
small to affect the results, so the ratio is usually stated as
$2n_1/c$. These experiments are frequently used to support the
Minkowski formulation, but we can see that the present treatment
also predicts the measured results. As this treatment is for
non-dispersive media, it does not say whether the term ${c}/{n_1}$
in this expression is the group velocity or the phase velocity of
the wave in the medium. The experiments show that it is the phase
velocity \cite{jones78}.

In a recent paper \cite{mansuripur07a}, Mansuripur treated the
problem of radiation pressure on mirrors immersed in linear
dielectric and non-magnetic media using a model of a medium with
imaginary refraction index to describe the mirrors. He considered
the case where the complex reflection coefficient of the mirror is
$\mathrm{{e}}^{i\phi}$ and calculated the Lorentz force on the
electric currents of the mirror. In the present treatment of the
problem of the reflection of an electromagnetic pulse by a
non-magnetic mirror with this complex reflection coefficient
immersed in a linear dielectric and magnetic medium, the momentum
transfer to the mirror can be calculated using Eqs. (\ref{pi pr
pt}), (\ref{p linha 1}), (\ref{p linha 3}) and (\ref{p linha 4}) as
$\mathbf{P}_{\mathrm{mir}}=\mathbf{P}_{\mathrm{i}}-\mathbf{P}_\mathrm{r}-\mathbf{P}'_1-\mathbf{P}'_3-\mathbf{P}'_4$
with $r=\mathrm{{e}}^{i\phi}$ and $\chi_{\mathrm{m2}}=0$. Retaining
terms up to the first power in $\chi_{\mathrm{m1}}$, we obtain
\begin{equation}\nonumber
    \mathbf{P}_{\mathrm{mir}}\simeq2\left[1+(\chi_{\mathrm{e1}}-\chi_{\mathrm{m1}}-\chi_{\mathrm{e1}}\chi_{\mathrm{m1}})\sin^2\left(\frac{\phi}{2}\right)\right]\mathbf{P}_0\;,
\end{equation} which is compatible with the result reported by
Mansuripur when $\chi_{\mathrm{m1}}=0$ and can be experimentally
tested. As I rely only on the properties of the linear medium and no
particular model to describe the mirror is adopted, this treatment
is more general than that of Mansuripur.

\section{Momentum transfer to an antireflection
layer}\label{sec:antirref}

In the next example, suppose that we have an antireflection layer
between media 1 and 2 consisted of a material with electric
susceptibility $\chi_\mathrm{e}'$ obeying
$(1+\chi_\mathrm{e}')=\sqrt{(1+\chi_{\mathrm{e1}})(1+\chi_{\mathrm{e2}})}$,
with magnetic susceptibility $\chi_\mathrm{m}'$ obeying
$(1+\chi_\mathrm{m}')=\sqrt{(1+\chi_{\mathrm{m1}})(1+\chi_{\mathrm{m2}})}$
and thickness ${\lambda'}/{4}$, $\lambda'$ being the wavelength of
the central frequency of the pulse in this medium. If we have an
almost monochromatic incident pulse in medium 1 propagating towards
the interface, it will be almost completely transmitted to medium 2.
The initial and final situations are illustrated in Fig. \ref{fig:
ant ref}. If the electric field of the incident pulse
$\mathbf{E}_\mathrm{i}$ is described by Eq. (\ref{pulso}), the
electric field of the transmitted one will be \cite{born}
\begin{eqnarray}\nonumber
\mathbf{E}_2(z,t)&=&\left[\frac{(1+\chi_{\mathrm{e1}})(1+\chi_{\mathrm{m2}})}{(1+\chi_{\mathrm{e2}})(1+\chi_{\mathrm{m1}})}\right]^{1/4}\times\\&&\times\frac{1}{\sqrt{2\pi}}\int_{-\infty}^{+\infty}
\mathrm{d}\omega
\tilde{E}(\omega)\mathrm{e}^{i\left(\frac{n_2\omega}{c}z-\omega
    t\right)}\mathbf{\hat{x}}\;.
\end{eqnarray}

\begin{figure}
  \begin{center}
  \includegraphics[width=6cm]{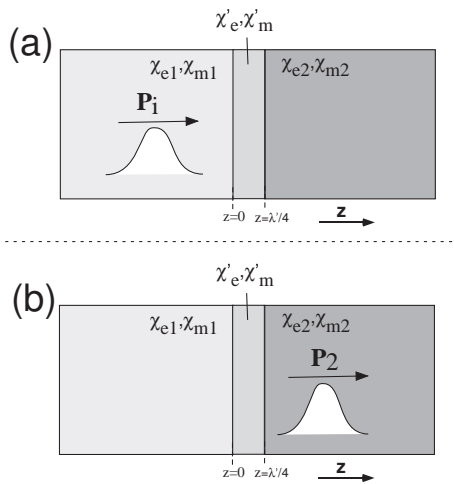}\\
  \caption{(a) A pulse with total momentum $\mathbf{P}_\mathrm{i}$ propagates in medium 1 with electric and magnetic susceptibilities
  $\chi_{\mathrm{e1}}$ and $\chi_{\mathrm{m1}}$
  towards  medium 2 with electric and magnetic susceptibilities
  $\chi_{\mathrm{e2}}$ and $\chi_{\mathrm{m2}}$. There is an antireflection coating layer between media 1 and 2
  consisted of a material with electric and magnetic
  susceptibilities
  $\chi'_{e}$ and $\chi'_{m}$ such that $(1+\chi_\mathrm{e}')=\sqrt{(1+\chi_{\mathrm{e1}})(1+\chi_{\mathrm{e2}})}$ and
  $(1+\chi_\mathrm{m}')=\sqrt{(1+\chi_{\mathrm{m1}})(1+\chi_{\mathrm{m2}})}$. The
thickness of the layer is ${\lambda'}/{4}$, $\lambda'$ being the
wavelength of the central frequency of the pulse in this medium.
Despite the figure, it is assumed that the pulse is much larger than
the layer. (b) The pulse was totally transmitted to medium 2 and has
total momentum $\mathbf{P}_2$.}\label{fig: ant ref}\end{center}
\end{figure}

The total momentum of the incident pulse $\mathbf{P}_\mathrm{i}$ and
of the transmitted pulse $\mathbf{P}_2$ can be written as
\begin{eqnarray}\label{p1 p2}
    \mathbf{P}_\mathrm{i}&=&\left(1+\frac{\chi_{\mathrm{e1}}+\chi_{\mathrm{m1}}+\chi_{\mathrm{e1}}\chi_{\mathrm{m1}}}{2}\right)\mathbf{P}_0\;\;,\\\nonumber\label{p1 p2 b}
    \mathbf{P}_2&=&\left[\frac{(1+\chi_{\mathrm{e1}})(1+\chi_{\mathrm{m2}})}{(1+\chi_{\mathrm{e2}})(1+\chi_{\mathrm{m1}})}\right]^{1/2}\times\\&&\times
    \left(1+\frac{\chi_{\mathrm{e2}}+\chi_{\mathrm{m2}}+\chi_{\mathrm{e2}}\chi_{\mathrm{m2}}}{2}\right)\mathbf{P}_0\;,
\end{eqnarray} with $\mathbf{P}_0$ given by Eq. (\ref{p0}). Since
the momentum of the initial and final pulses are different, there
must be a momentum transfer to the antireflection layer in order
that we have momentum conservation. Now I will show that the use of
the force density of Eq. (\ref{force density}) acting in the
antireflection layer guarantees momentum conservation in the
process. Let us call $\mathbf{E}'$ and $\mathbf{B}'$ the electric
and magnetic fields in the region of the layer. The boundary
conditions impose \begin{eqnarray}\nonumber
&&\mathbf{E}'|_{z=0}=\mathbf{E}_{1}|_{z=0}\;,\;
\mathbf{E}'|_{z=\lambda'/4}=\mathbf{E}_{2}|_{z=\lambda'/4}\;,\;\\
&&\frac{\mathbf{B}'|_{z=0}}{1+\chi_\mathrm{m}'}=\frac{\mathbf{B}_{1}|_{z=0}}{1+\chi_{\mathrm{m1}}}\;,\;
\frac{\mathbf{B}'|_{z=\lambda'/4}}{1+\chi_\mathrm{m}'}=\frac{\mathbf{B}_{2}|_{z=\lambda'/4}}{1+\chi_{\mathrm{m2}}}\;.
\end{eqnarray} Writing $\mathbf{f}_3$ from Eq. (\ref{force density})
as in Eq. (\ref{f3}) and integrating the force density $\mathbf{f}$
on time and in the volume of the layer, we obtain the momentum
transfer to the layer: \begin{eqnarray}\label{pa}
   \mathbf{P}''_\mathrm{a}&=&\int \mathrm{d}x \int \mathrm{d}y\int^{\lambda'/4}_0 \mathrm{d}z\int_{-\infty}^{+\infty} \mathrm{d}t\;
    \mathbf{f}\\\nonumber
    &=&\frac{\chi_\mathrm{e}'+\chi_\mathrm{m}'+\chi_\mathrm{e}'\chi_\mathrm{m}'}{2}\left[\sqrt{\frac{(1+\chi_{\mathrm{e1}})(1+\chi_{\mathrm{m2}})}{(1+\chi_{\mathrm{e2}})(1+\chi_{\mathrm{m1}})}}-1\right]\mathbf{P}_0.
\end{eqnarray}

We must also consider the momentum transfers to the interfaces
between the layer and the mediums 1 and 2 due the discontinuities of
the magnetization $\mathbf{M}$. Calling $\mathbf{P}''_\mathrm{b}$
the momentum transfer in $z=0$ and $\mathbf{P}''_\mathrm{c}$ the
momentum transfer in $z=\lambda'/4$ and repeating the treatment of
Sec. \ref{sec:int diel}, we obtain \begin{eqnarray}\label{pb pc}
  \mathbf{P}''_\mathrm{b} &=&  \frac{(\chi_{\mathrm{m1}}-\chi'_{m})(2+\chi_{\mathrm{m1}}+\chi'_{m})(1+\chi_{\mathrm{e1}})}{2(1+\chi_{\mathrm{m1}})}\mathbf{P}_0,
  \\\nonumber \label{pb pc b}
  \mathbf{P}''_\mathrm{c} &=&
  \frac{(\chi'_{m}-\chi_{\mathrm{m2}})(2+\chi'_{m}+\chi_{\mathrm{m2}})(1+\chi_{\mathrm{e2}})}{2(1+\chi_{\mathrm{m2}})}\times\\&&\times\sqrt{\frac{(1+\chi_{\mathrm{e1}})(1+\chi_{\mathrm{m2}})}{(1+\chi_{\mathrm{e2}})(1+\chi_{\mathrm{m1}})}}\mathbf{P}_0\;.
\end{eqnarray}

The total momentum transfer to the antireflection layer due to the
passage of the pulse is
$\mathbf{P}''=\mathbf{P}''_\mathrm{a}+\mathbf{P}''_\mathrm{b}+\mathbf{P}''_\mathrm{c}$.
Using Eqs. (\ref{p1 p2}), (\ref{p1 p2 b}), (\ref{pa}), (\ref{pb pc})
and (\ref{pb pc b}), we can see that \begin{equation}
    \mathbf{P}_2+\mathbf{P}''=\mathbf{P}_\mathrm{i}
\end{equation} and momentum is conserved in the process. Again, we
see the consistency of the proposed division of the momentum of the
wave.

\section{Einstein box theories}\label{sec:einstein}

As a last example, we can see whether the present treatment agrees
with Einstein's theory of relativity testing a \emph{gedanken}
experiment of the kind ``Einstein box theories'' in which a single
photon in free space with energy $\hbar \omega$ and momentum
$\mathbf{P}_0=({\hbar \omega}/{c})\mathbf{\hat{z}}$ is transmitted
through a transparent slab with electric and magnetic
susceptibilities $\chi_{\mathrm{e2}}$ and $\chi_{\mathrm{m2}}$,
refraction index
$n_2=\sqrt{(1+\chi_{\mathrm{e2}})(1+\chi_{\mathrm{m2}})}$, length
$L$, mass $M$ and antireflection layers in both sides. Due to
propagation in the medium, the photon suffers a spacial delay
$(n_2-1)L$ in relation to propagation in vacuum. According to the
theory of relativity, the velocity of the center of mass-energy of
the system must not change due to the passage of the photon through
the slab, so the slab must suffer a displacement $\Delta z$ such
that \cite{frisch65,brevik81,loudon04} \begin{equation}\label{centro
de energia}
    Mc^2\Delta z=\hbar\omega(n_2-1)L\;.
\end{equation} The use of the Abraham momentum density as the
electromagnetic part of the total momentum density gives directly
the correct displacement of the slab, so this \emph{gedanken}
experiment is frequently used to support the Abraham formulation.
Now I will show that the present treatment also gives the correct
displacement in a direct way. The mechanical momentum of the slab
$\mathbf{P}_{\mathrm{slab}}$ during the passage of the photon has 2
contributions. The first is the momentum transferred to the first
antireflection layer. The second is the material part of the
momentum of the photon in the slab. Using Eqs. (\ref{pmat}),
 (\ref{p1 p2 b}), (\ref{pa}),(\ref{pb pc}) and
(\ref{pb pc b}) with $\chi_{\mathrm{m1}}=\chi_{\mathrm{e1}}=0$, we
find that the total momentum of the slab is \begin{equation}
    \mathbf{P}_{\mathrm{slab}}=\frac{n_2-1-\chi_{\mathrm{m2}}}{n_2}\frac{\hbar
\omega}{c}\mathbf{\hat{z}}\;. \end{equation} After the passage of
the pulse, the momentum of the slab comes back to zero. Part of this
momentum is in the form of hidden momentum
\cite{penfield,schockley67,vaidman90}, a relativistic effect that is
not associated with the movement of the slab. A magnetic dipole
$\mathbf{m}$ in an uniform electric field $\mathbf{E}_0$ has a
hidden momentum $\mathbf{m}\times \mathbf{E}_0$. So the density of
hidden momentum of an electromagnetic wave in a linear medium is
given by $\mathbf{M}\times \mathbf{E}$. Integrating this density in
volume, the total hidden momentum of the pulse in the slab is given
by $\mathbf{P}_{\mathrm{hid}}=-{\chi_{\mathrm{m2}}\hbar
\omega}/({n_2}{c})\mathbf{\hat{z}}$. To find the velocity of the
slab, we must subtract the hidden momentum from its total momentum
and divide by its mass. As the pulse takes a time ${n_2L}/{c}$ to
pass through the slab, the displacement of the slab is
\begin{equation}\nonumber
    \Delta
    z=\frac{|\mathbf{P}_{\mathrm{slab}}-\mathbf{P}_{\mathrm{hid}}|}{M}\frac{n_2L}{c}=\frac{\hbar\omega(n_2-1)L}{Mc^2}\;,
\end{equation} in agreement with Eq. (\ref{centro de energia}). One
more time we see the consistency of the proposed division of the
momentum of the wave into electromagnetic and material parts.

\section{Conclusions}\label{conc}

In summary, it was shown that the momentum density of
electromagnetic waves in linear non-absorptive and non-dispersive
dielectric and magnetic media can be naturally and consistently
divided into an electromagnetic part
$\varepsilon_0\mathbf{E}\times\mathbf{B}$, which has the same form
independently of the medium, and a material part that can be
obtained directly using the Lorentz force. This division was shown
to be consistent with momentum conservation in various circumstances
and with the ``Einstein box theories''. I believe that it may be
possible to extend this division to all kinds of media.

I also calculated permanent transfers of momentum to the media due
to the passage of electromagnetic pulses that were missing in
previous treatments \cite{loudon02,loudon03,mansuripur04}, showed
the compatibility of the division with the experiments of Jones
\emph{et al.} \cite{jones54,jones78} and generalized the treatment
of radiation pressure on submerged mirrors \cite{mansuripur07a},
which can be submitted to experimental verification.

\acknowledgments

The author acknowledges Carlos H. Monken and J\'ulia E. Parreira for
useful discussions and C\'elio Zuppo for revising the manuscript.
This work is supported by the Brazilian agency CNPq.


\end{document}